\documentclass[prl,aps,floatfix,twocolumn,preprintnumbers,superscriptaddress, showpacs]{revtex4-1}
\usepackage{graphicx}
\usepackage{dcolumn}
\usepackage{bm}
\usepackage{amssymb}
\usepackage{pifont}
\usepackage{amsmath}
\usepackage{times}
\usepackage[nooneline]{subfigure}

\begin{document}


\title{Fluctuations and scaling in creep deformation}
\author{Jari Rosti} \email{jro@fyslab.hut.fi}
\affiliation{Department of Applied Physics, Aalto University, PO Box
14100, Aalto 00076, Finland}

\author{Juha Koivisto} \email{jko@fyslab.hut.fi}
\affiliation{Department of Applied Physics, Aalto University, PO Box
14100, Aalto 00076, Finland}

\author{Lasse Laurson} \email{lasse.laurson@gmail.com}

\affiliation{ISI Foundation, Viale S. Severo 65, 10133 Torino, Italy}

\author{Mikko J.\ Alava} \email{mikko.alava@tkk.fi}

\affiliation{Department of Applied Physics, Aalto University, PO Box
14100, Aalto 00076, Finland}

\date{\today}


\begin{abstract}
The spatial fluctuations of deformation are studied in creep in the
Andrade's power-law and the logarithmic phases, using paper samples.
Measurements by the Digital Image Correlation technique show that
the relative strength of the strain rate fluctuations increases with
time, in both creep regimes. In the Andrade
creep phase characterized by a power law decay of the strain rate
$\epsilon_t \sim t^{-\theta}$, with $\theta \approx 0.7$, the
fluctuations obey $\Delta \epsilon_t \sim t^{-\gamma}$, with $\gamma
\approx 0.5$. The local deformation follows a data
collapse appropriate for an absorbing state/depinning transition.
 Similar behavior is found in a
crystal plasticity model, with a jamming or yielding phase
transition.
\end{abstract}

\maketitle

The creep of a sample under a constant load exhibits often three
rheological phases in materials from metals and alloys with a
crystalline structure to rocks and composites: the initial power-law
decrease of the strain rate or primary creep, quasi-stationary
secondary creep, and finally tertiary creep approaching final
failure. It is important to understand such empirical laws of
scaling in material deformation. The Andrade creep law, stating that
the deformation rate decays in time as $\frac{d \epsilon(t)}{dt}
\equiv \epsilon_t \sim t^{-\theta}$, with $\theta \approx 2/3$,
originates already from 1910 
\cite{andrade}. The phenomenological creep laws \cite{louchet}  are
valid for many materials from metals to polymeric materials to
composites like ordinary paper \cite{coffin} to rocks
\cite{nabarro}. In the case of metals the dynamics is related to
dislocations and the structures they form. It is intriguing why
there should be such universality, which seems to be
"coarse-grained", not dependent on the microscopic details.

In this vein, collective phenomena as fluctuations and avalanches
\cite{sethna} have been recently studied in crystal plasticity
\cite{miguel,weiss,jakobsen,dimiduk,csikor}, viscoelastic fluids and
networks \cite{coussot,liu2,bocquet}, granular materials and
amorphous glasses \cite{maloney}  and in fracture
\cite{nechad,alava,kun,bonamy}. The dynamics of simple dislocation
assemblies exhibits a phase transition at a critical external shear
stress value $\sigma_c$, similar to the jamming one
seen in granular media and glasses \cite{liu,goyon,keys}, with the
order parameter (the deformation rate) exhibiting a power-law decay
in time close to the transition, as the Andrade law \cite{miguel2,lasse}.

To explore the physics of creep and to see if it could be related to
such collective behavior, we measure local creep rates in primary
creep. The main question is: what kind of fluctuations does the
creep show, and how do they scale with time?  In non-crystalline
materials strain variations implies a localization of deformation
which is unrelated to dislocation activity. So, another issue is:
what kind of universality exists in creep deformation and in local
creep strain rates? By digital image correlation analysis, we show
that the spatial fluctuations of creep strain rates follow a scaling
law during the Andrade creep. The ratio of fluctuations to mean rate
increases with time. This feature persists in the logarithmic creep
regime, where $\epsilon(t) \sim \log(t)$, or $\epsilon_t \sim
t^{-1}$. The critical properties of the power-law creep are tested
by applying the theory of non-equilibrium phase transitions. We find
that spatial creep deformation statistics follows its own scaling
and a scaling function. Similar behavior is encountered in a simple
discrete dislocation or crystal plasticity model, which exhibits
Andrade creep for applied stress values close to the critical stress
$\sigma_c$ of the dislocation jamming transition. This stress
separates an active phase with a constant average creep rate 
from a low-stress phase where the deformation rate dies out
after a transient \cite{miguel2}. The temperature enters indirectly
only via the dislocation mobility. The most natural explanation is
that Andrade creep derives from a non-equilibrium phase transition
between "jammed", or frozen, and "flowing", or active, states
\cite{liu}. Usual materials science constitutive equations and creep
laws are based on phenomenological equations, usually related to an
thermally activated process \cite{lin}. They may fail to cope with
correlated, intermittent, dynamics.

To induce creep in experiments, we applied a constant stress to a
sheet of paper in a tensile-test geometry (Fig. 1a). Tests were made
in humidity and temperature controlled conditions, $33\pm1$ \% of
relative humidity and $33\pm1$ $^o$C of temperature, with the typical
applied loads around 40 N, corresponding to a stress of 13 MPa. We used
ordinary office paper, with 30~$\times$~100~mm$^2$ samples loaded in
the cross-machine direction. The sample is imaged with SensiCam
370KL0562 digital camera with a low thermal noise ratio, at 12 bit
gray scale resolution.
The exposure time in measurements was 200~ms, and the
sampling frequency of the digital imaging was $0.1$~Hz.

In sample scale images, a printed speckle pattern (Fig. 1b) was used
due to its sharp contrast and translational invariance. A set of
experiments with larger magnification and smaller load was also made
with an image size of 3x4 mm and using specially prepared sheets,
with 5 \% of the fibres colored for contrast. The larger
magnification and a smaller load make it possible to follow the
entire experiment with Digital Image Correlation (DIC), beyond the
primary creep regime.

Experimentally, a time-series of displacement fields is derived
using the DIC-technique \cite{sutton,kybic,hild} (Fig. 1a). The algorithm describes both the image and the
deformation using the B-spline model. The algorithm finds the
deformation function by using the multiresolution approach for the
minimization, so that the image and the deformation model is refined
every time the convergence is reached. The criteria for the
convergence is the sum of squared differences (SSD), i.e. the
deformation function is applied to the original image and the SSD is
computed against the deformed image. The algorithm is described in Ref.~\cite{kybic2}. A
discrete picture, in space and time, ensues of the local creep
strain rates \cite{long}. An example (Fig. 1b) shows the variations
of the rates for one $\Delta t$. For the spatial fluctuations of the
local strain rate $\epsilon_t^{i,j}$, we define an evenly spaced
grid on an image, which consists of 49$\times$78 points.  The
standard deviation of local strain rates $\Delta \epsilon_t$ is
computed over the grid values  on the lower half on the sample where
absolute displacements are large. This is further illustrated in
Figure 1a, which shows how the creep fluctuations are
measured from the experiment.

A feature visible in experiments is the disordered structure of paper,
two-dimensional over scales larger than the out-of-plane thickness \cite{alava2}.
One should note that the non-uniformity results in a locally varying
stress field, even in the
elastic deformation regime, in analogy to the internal
stresses of materials with dislocations.

The Andrade creep results are confronted with those of the
simulations of a two-dimensional discrete dislocation dynamics (DDD)
model \cite{giessen,miguel,miguel2}. The $N$
point-like dislocations with Burgers vectors ${\bf b}_n = s_n b {\bf
u}_x$ (with $s_n = \pm 1$ and $n=1 \dots N$) parallel to the glide
direction can be thought to represent a cross-section of a single
crystal with straight parallel edge dislocations and a single slip
geometry \cite{miguel}. The dislocations interact through their
anisotropic long-range shear stress fields, and perform over-damped
glide motion only driven by an applied external shear stress
$\sigma$ \cite{miguel,lasse,long}.
Fluctuations are studied in the vein illustrated by Figure 1a by
measuring the local strain rates
\begin{equation}
\epsilon_t^{i,j}(t) = \frac{b}{l^2}\sum_{\mathbf{r}_n \in \ \mathrm{box}\ i,j}
s_n b \left[\frac{x_n(t+0.5 \Delta t) - x_n(t-0.5 \Delta t)}{\Delta t}\right]
\end{equation}
over a time scale $\Delta t$ within sub-volumes of linear
size $l$, with $x_n(t)$ the position of the $n$th dislocation within
its glide plane.

Figure 2 depicts the Andrade creep results for the standard
deviation of the creep rate probability distributions. For both
paper samples and DDD simulations (performed at $\sigma=\sigma_c
\approx 0.025$, in dimensionless units \cite{long})
we observe a typical Andrade law of the form  $\epsilon_t \propto t^{-0.7}$
or an Andrade-exponent $\theta$ close to
2/3. The standard deviation 
follows also a power-law, and scales as
\begin{equation}
\Delta \epsilon_t \propto t^{-0.5},
\end{equation}
so an exponent $\gamma \sim 0.5$ is found. The "second" Andrade 
or fluctuation exponent $\gamma$ is slightly higher for experiments
(0.55), while in the DDD simulations a value of 0.5 is indicated.
Evidently this exponent is smaller than
that of the mean creep rate. Thus fluctuations increase in relative
terms, and creep becomes more intermittent with time. The accuracy
of the DIC method poses a limit to following the power-law of
Eq.~(1), apart from the maximum length of the power-law creep regime
obtainable by tuning the parameters. In
the DDD model
the scaling range is controlled by the proximity of the
jamming transition.

At lower stresses, logarithmic creep is found with no signature of the Andrade
phase. 
Figure 3a shows that a fluctuation scaling can be extracted in this
regime. Still, the fluctuation magnitude decays slower than the
creep rate. Their ratio might be related in each sample to the transition to tertiary
creep, which might indicate a zone-like localization of
subsequent deformation. The simultaneous measurement of Acoustic
Emission indicates no substantial damage accumulation or
micro-cracking until close to the transition to tertiary creep.
Thus DIC measures the statistics of viscoplastic
deformation, and not deformations related to eventual brittle
fracture. Microscopically in paper, the fibers and the network
structure can both deform irreversibly. For the sample-to-sample
variations, we find here a Monkman-Grant relation
\cite{monkman,nechad} between the time at which a minimum creep rate
occurs in a sample $t_m$ and the final failure time $t_c$, $t_m
\propto 0.83\, t_c$. The relevance of fluctuations for the
Monkman-Grant law would be interesting to understand.

The theoretical interpretation of the scaling of the creep 
fluctuations can be analyzed within and with the aid of the DDD
model. One question is how
this "second Andrade-scaling" is related to the
usual primary creep exponent? Assuming intermittent, localized creep
activity the active fraction of the regions of the specimen account
for the creep strain measured in a time step. This argument leads
to a self-consistency condition between the "number" of such active
regions as a function of time, and the typical time-dependent creep
strain rate in each such region \cite{long}. If the statistics 
would originate from the decay of the fraction of
localized, active regions, themselves deforming at a rate
independent of time, it would indicate a $\gamma=1/3$, half of the
Andrade $\theta$, instead
of the measured 0.5 $\dots$ 0.55. This line of reasoning is related
to classical, dislocation-based theory arguments of Andrade
creep \cite{mott,cottrell}.

The nature of the creep dynamics seems to exhibit universality
typical of phenomena arising from phase transitions. If a wide
variety of materials follows such behavior, it calls for a continuum
description, independent of microscopic details. Statistical physics
offers concepts such as depinning transitions of elastic manifolds
in random media. These have been applied to coarse-grained
dislocation plasticity \cite{zaiser}, and are examples 
of absorbing-state phase transitions.
Similarly to what is seen in the DDD model, such theories present or
predict a power-law relaxation of the order parameter close to the
critical point. The jamming/absorbing state transition scenario can
be further tested by the ansatz that the creep rate is the "order
parameter" of the phase transition. One collapses the total
deformation histograms, $P(\epsilon,t)$ (scaling out the average
deformation at time $t$) as one should do 
\cite{dickman} for the integrated order parameter. Both the DIC
(Figure 3b) and DDD data (\cite{long}) exhibit a reasonable 
data collapse, indicating critical behaviour, close to such a 
phase transition. Further evidence for the origin of the fluctuations in
collective phenomena can be sought in the correlation functions
$C(x-x',t-t')$ of strain fields, calculated as for an "interface
field" $\epsilon(x,t)$. Preliminary results imply, that the temporal
scaling yields the Andrade exponent \cite{long}.
The fact that the jamming transition picture works for the
experiments seems as such surprising for a non-crystalline material,
though a possible universality of the Andrade creep would argue for
the opposite.

It remains to extend the measurements to study the small scale
spatiotemporal features, such as
localized events \cite{events}. The creep fluctuations could be
investigated in other systems such as two- and
three-dimensional colloidal crystals or glasses \cite{video,schall}.
Future research directions include considering the
fluctuations for various load histories, an example being fatigue,
and the role of "aging" and sample-relaxation.

Local creep rates exhibit a fluctuation scaling,
both in a disordered, viscoplastic material and
in a model of dislocation plasticity. This
presents a challenge: what kind of coarse-grained theory
for the deformation field would reproduce those signatures?
Such fluctuations are beyond classical,
phenomenological creep models. The phase transition scenario of
Andrade creep relegates the temperature to a secondary role,
in the DDD model via the viscous dynamics of the
dislocations. In depinning, for stresses below the critical one, 
thermal fluctuations induce a transient towards a thermally activated steady-state
creep  \cite{kolton}, which could relate to the logarithmic behavior
in experiments.


Non-equilibrium scenarios include
various models for the plasticity of amorphous materials
\cite{bocquet,falk,maloney,vande}
where the presence of localized events
has been recognized e.g. in the so-called Shear Transformation Zone
theory. To summarize, the main issue is that creep deformation
exhibits spatial fluctuations which scale in time, indicating that
the different phenomenological creep regimes could be understood in
terms of phase transitions.


\begin{figure}
\centerline{\includegraphics[width=75mm,type=eps,ext=.eps,read=.eps]{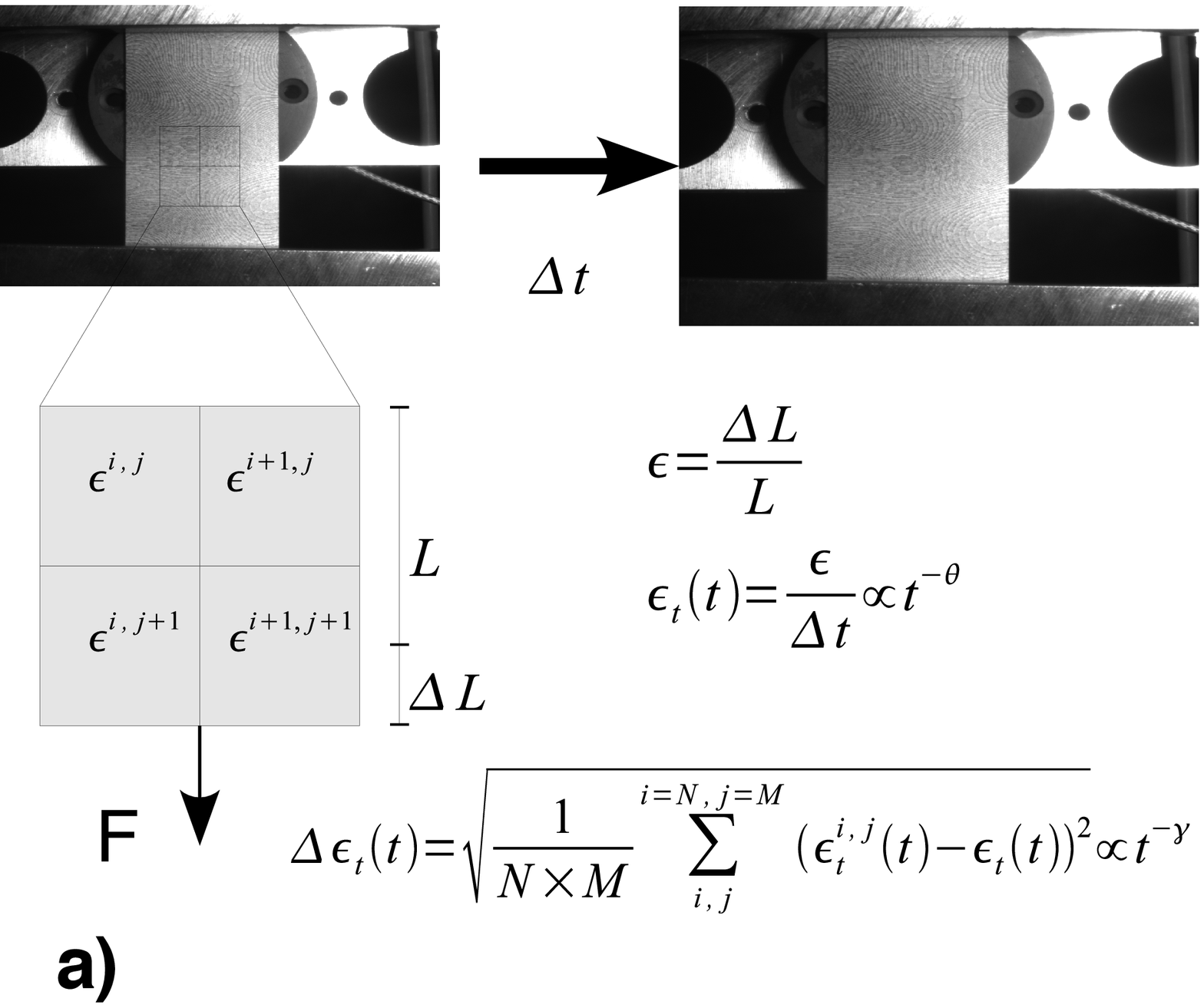}}
\centerline{\includegraphics[width=75mm,type=eps,ext=.eps,read=.eps]{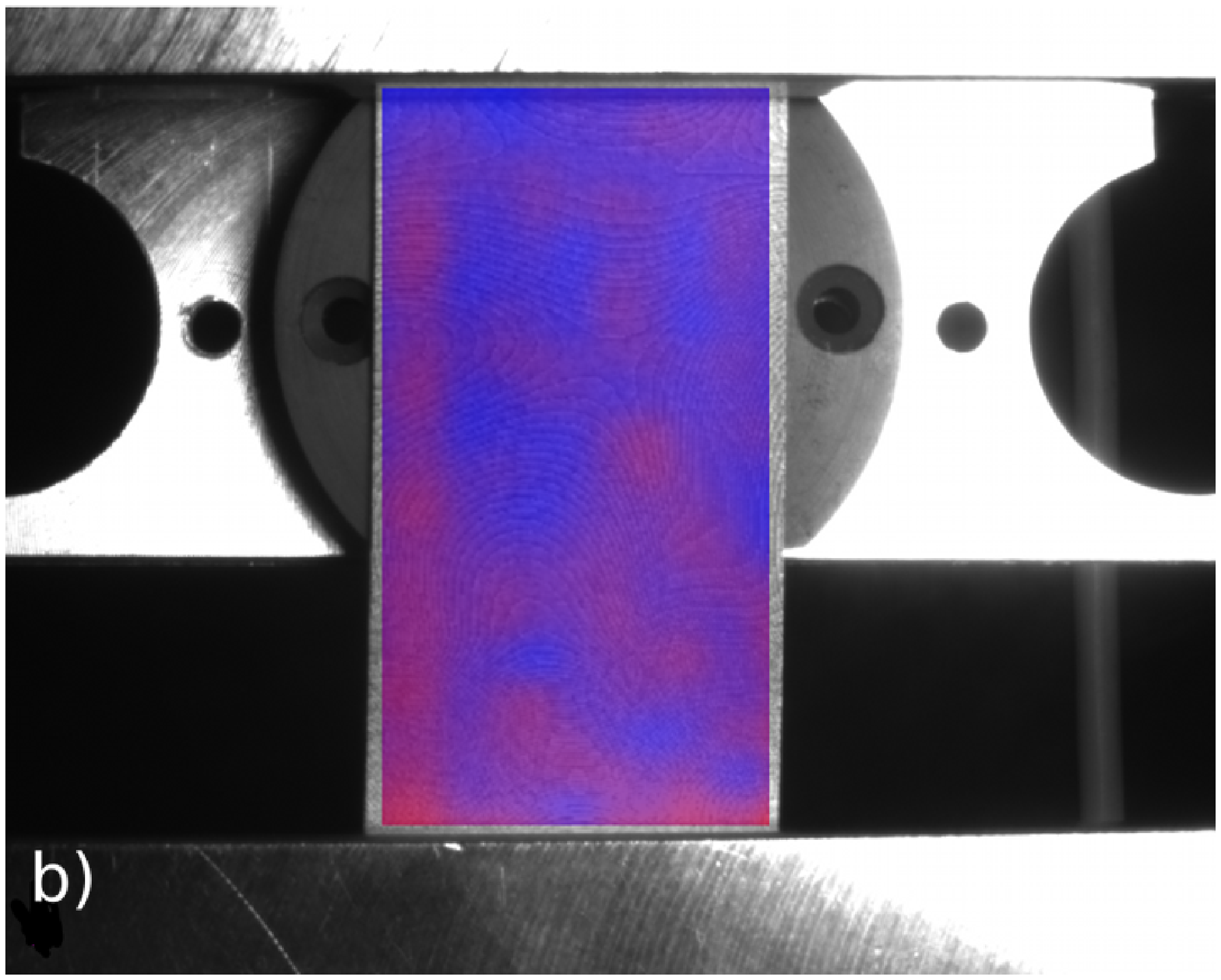}}
\caption{The experimental scenario. From a pair of digital images at
a time-interval $\Delta t$ the local deformations are extracted at a grid of
$N \times M$ points, with $N=49$ and $M=39$ on the lower half plane of the sample.
The strain fluctuations are measured
via the time-dependent standard deviation, and compared to the mean creep rate,
here in primary/Andrade creep. In the DDD simulations, the system 
is similarly divided into boxes of linear size $l$, with
local strain rates $\epsilon_t^{i,j}$. b) A digital
image of a paper sample on a scale of 40 mm$^2$. Superposed is a typical deformation grid for a
time difference $\Delta t$ of 10 seconds. The color scale indicates the degree
of local creep deformation (blue: small, red: large). In the
background: the experimental setup. The visible speckle pattern is
printed and designed to have a structure and contrast
appropriate for the DIC method. } \label{fig1}
\end{figure}
\begin{figure}
\centerline{\includegraphics[width=75mm,type=eps,ext=.eps,read=.eps]{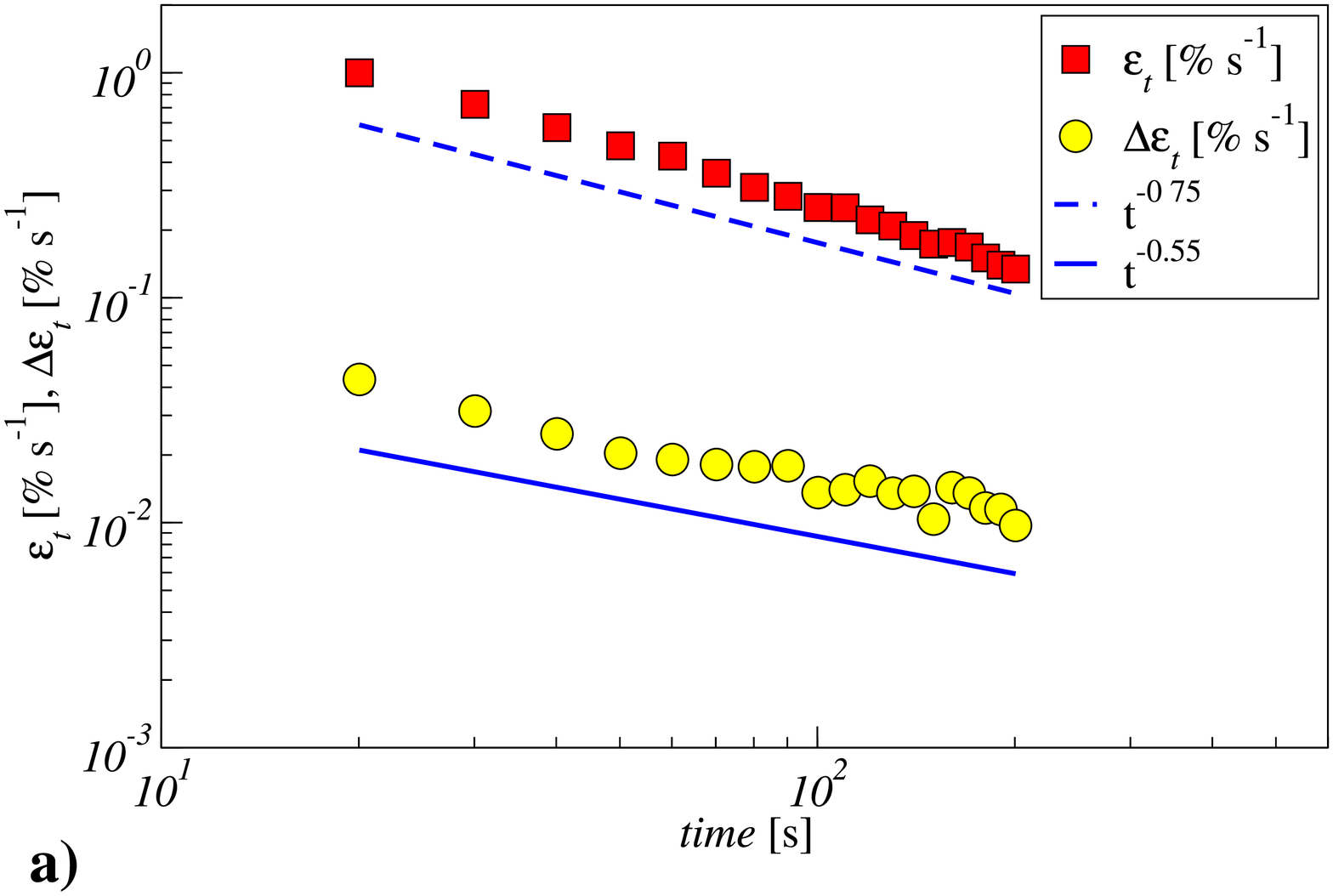}}
\centerline{\includegraphics[width=75mm,type=eps,ext=.eps,read=.eps]{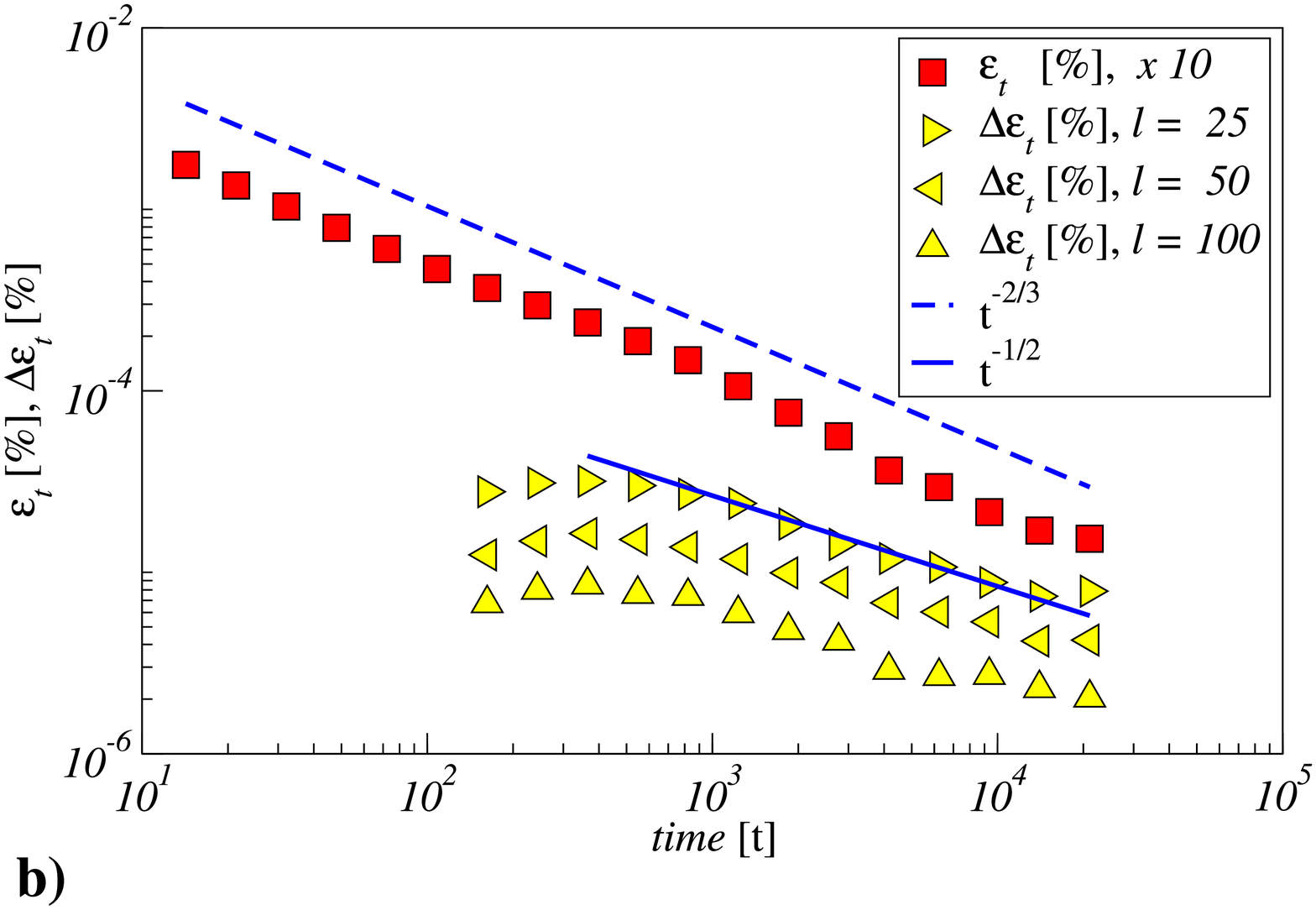}}
\caption{(Color online) Average creep rate and the standard deviation of the local
creep rates. a) Typical experimental data. The sample
average lifetime is 800-1600 seconds. The primary-to-secondary creep
transition takes place continuously at 150-200 seconds. 16
experiments are used. The effective Andrade
exponent is about 0.7 in the range 10 to 50 seconds, and becomes
close to one (signalling secondary creep) above 150
seconds.  b) The Andrade's law and the
fluctuation scaling for the DDD model for three different box sizes
to compute the local rates, at $\sigma \approx \sigma_c$. The early-time cross-over in the
fluctuations is due to computation time-interval of $\Delta t=100$.} \label{fig2}
\end{figure}
\begin{figure}
\centerline{\includegraphics[width=75mm,type=eps,ext=.eps,read=.eps]{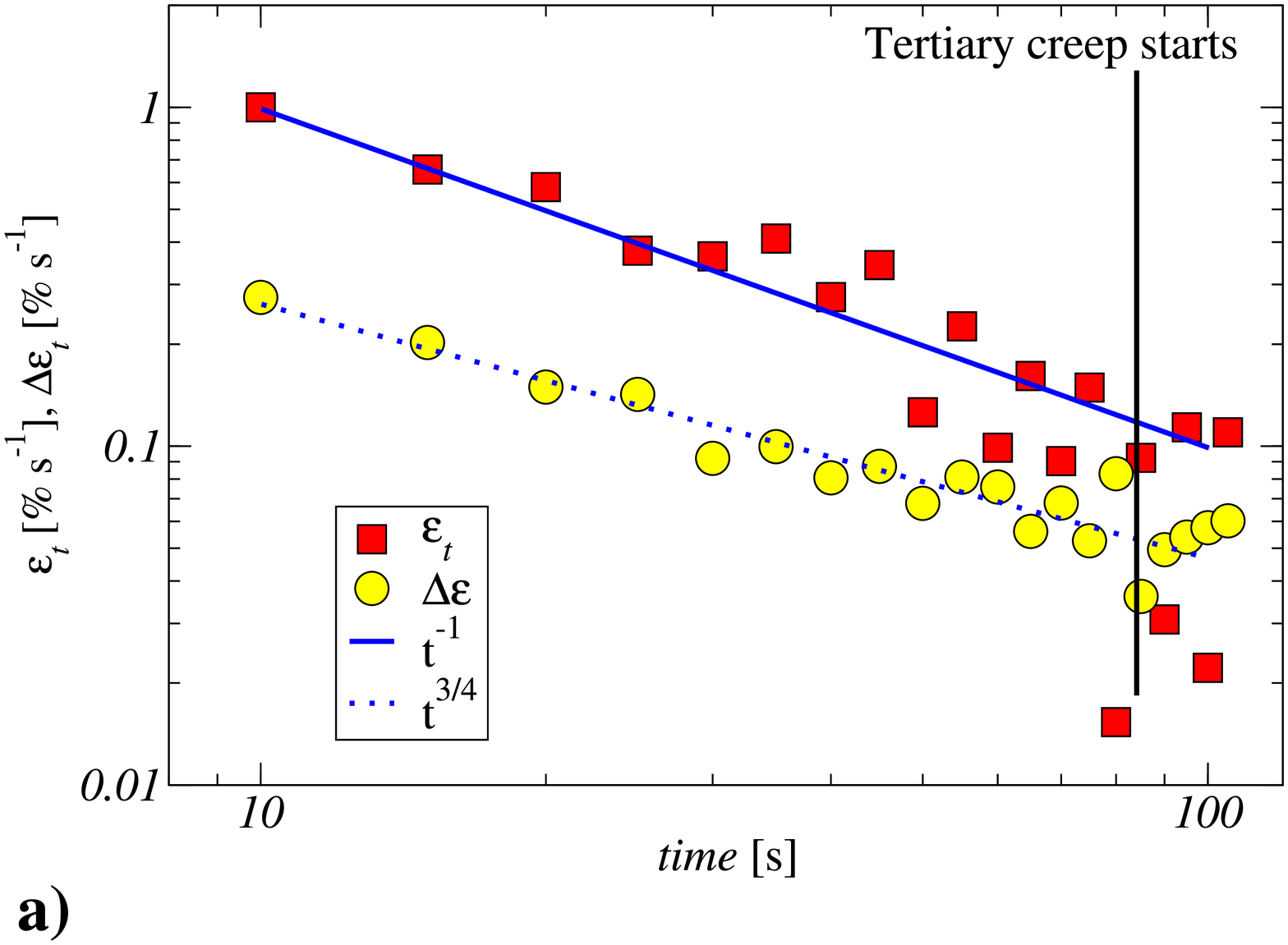}}
\centerline{\includegraphics[width=75mm,type=eps,ext=.eps,read=.eps]{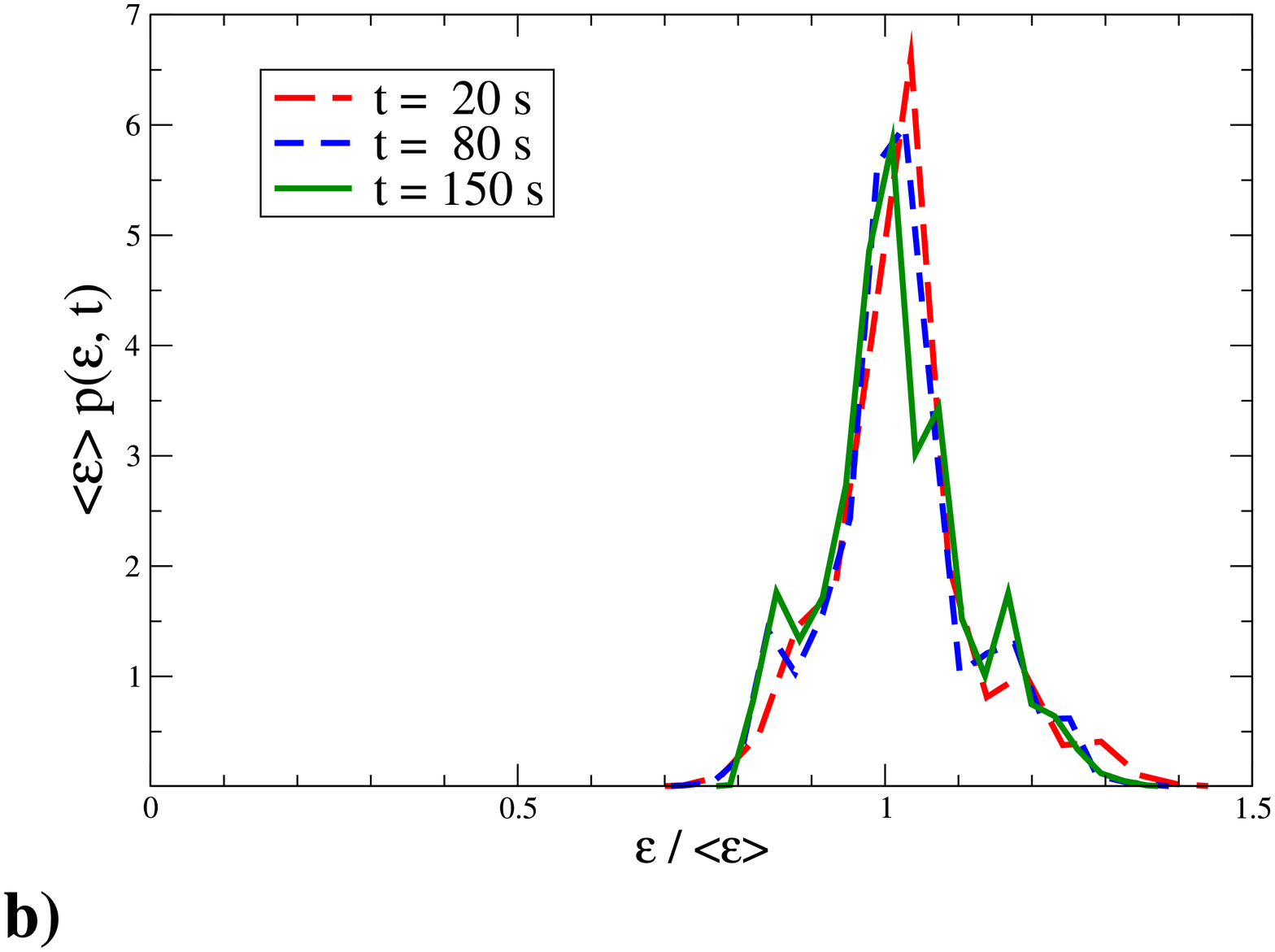}}
\caption{(Color online) a) Experimental results from logarithmic creep using a smaller
imaging area, 5 second image intervals, and a smaller load, showing the average creep rate
and the fluctuations. The former decays faster,  and at the end
these two become comparable, on the observation scale.
The data is an average over 9 experiments. b) Data collapse of the primary creep strain
distribution $P(\epsilon,t)$ from the experiments. The local
deformation distributions can be collapsed using the creep strain
as the scaling parameter.} \label{fig3}
\end{figure}


\acknowledgments We acknowledge the support of the Academy of
Finland via the Center of Excellence program and a post-doctoral
grant, and the European Commission NEST Pathfinder programme TRIGS
under contract NEST-2005-PATH-COM-043386.

\bibliographystyle{prsty}

\end{document}